\documentclass[prd,aps,twocolumn,superscriptaddress,letterpaper,longbibliography]{revtex4-1}
\usepackage{amssymb}
\usepackage[english]{babel}
\usepackage{mathtools}
\usepackage[pdftex]{graphicx}
\usepackage[usenames, dvipsnames, svgnames, table]{xcolor}
\usepackage[colorlinks=true, citecolor=Blue,
            linkcolor=BrickRed, urlcolor=ForestGreen]{hyperref}
\usepackage{txfonts}
\usepackage{mathrsfs}
\usepackage{bm}

\begin{document}


\newcommand{\be}{\begin{equation}}
\newcommand{\ee}{\end{equation}}
\newcommand{\R}[1]{\textcolor{red}{#1}}
\newcommand{\B}[1]{\textcolor{blue}{#1}}
\newcommand{\fixme}[1]{\textcolor{orange}{#1}}

\def\earm{\epsilon_{\rm arm}}
\def\esrc{\epsilon_{\rm src}}
\def\eext{\epsilon_{\rm ext}}
\def\eint{\epsilon_{\rm int}}
\def\eout{\epsilon_{\rm out}}

\def\Titm{T_{\rm itm}}
\def\Tsrc{T_{\rm src}}

\def\Mrot{{\bf M}_{\rm rot}}
\def\Msqz{{\bf M}_{\rm sqz}}
\def\Mopt{{\bf M}_{\rm opt}}
\def\Mio{{\bf M}_{\rm io}}
\def\Mc{{\bf M}_{\rm c}}


\title{Broadband Quantum Noise Reduction in Future Long Baseline Gravitational-wave Detectors via EPR Entanglement}

\author{Jacob L. Beckey}
\affiliation{School of Physics and Astronomy, University of Birmingham, Birmingham, B15 2TT, United Kingdom
} 
\author{Yiqiu Ma}
\affiliation{Center for Gravitational Experiment, School of Physics, Huazhong University of Science and Technology, Wuhan 430074, China}
\affiliation{Theoretical astrophysics 350-17, Californian Institute of Technology, Pasadena, Californian 91125, USA}

\author{Vincent Boyer}
\affiliation{School of Physics and Astronomy, University of Birmingham, Birmingham, B15 2TT, United Kingdom
} 

\author{Haixing Miao}
\affiliation{School of Physics and Astronomy and Institute of Gravitational Wave Astronomy,
University of Birmingham, Birmingham, B15 2TT, United Kingdom}


\begin{abstract}
Broadband quantum noise reduction can be achieved in gravitational wave detectors by injecting frequency dependent squeezed light into the the dark port of the interferometer. This frequency dependent squeezing can be generated by combining squeezed light with external filter cavities. However, in future long baseline interferometers (LBIs), the filter cavity required to achieve the broadband squeezing has a low bandwidth -- necessitating a very long cavity to mitigate the issue from optical loss. It has been shown recently that by taking advantage of Einstein-Podolsky-Rosen (EPR) entanglement in the squeezed light source, the interferometer can simultaneously act as a detector and a filter cavity. This is an attractive broadband squeezing scheme for LBIs because the length requirement for the filter cavity is naturally satisfied by the length of the interferometer arms. In this paper we present a systematic way of finding the working points for this broadband squeezing scheme in LBIs. We also show that in LBIs, the EPR scheme achieves nearly perfect ellipse rotation as compared to 4km interferometers which have appreciable error around the intermediate frequency. Finally, we show that an approximation for the opto-mechanical coupling constant in the 4km case can break down for longer baselines. These results are applicable to future detectors such as the 10km Einstein Telescope and the 40km Cosmic Explorer.
\end{abstract}

\maketitle


\section{Introduction}
Gravitational-wave (GW) detectors including LIGO and VIRGO, which recently made breakthrough discoveries are Michelson-type interferometers with km size arms\,\cite{FirstGW,GW2}. They are among the largest and most sensitive experiments humans have ever constructed. To push the limits of scientific discovery even further, larger, more sensitive instruments are already being planned. Two such detectors are the 10km Einstein Telescope (ET) \cite{ETDesign} and the 40km Cosmic Explorer \cite{Abbott_2017}. They differ from LIGO in many ways including scale and configuration, but for our purposes can be treated in a very similar way mathematically.

All ground-based GW detectors are plagued by various noise sources that result from the fact that they are on Earth (e.g. seismic activity). Once these and all other classical noise sources are suppressed, the sensitivity of GW detectors is ultimately limited by the quantum nature of light. The quantized electromagnetic field is analogous to a quantum harmonic oscillator (position and momentum of a mass are replaced by amplitude and phase quadrature of light). The uncertainty in the amplitude and phase quadrtaures (quantum fluctuations) limits the sensitivity of interferometric measurements.

\begin{figure}[b]
  \includegraphics[width=\columnwidth]{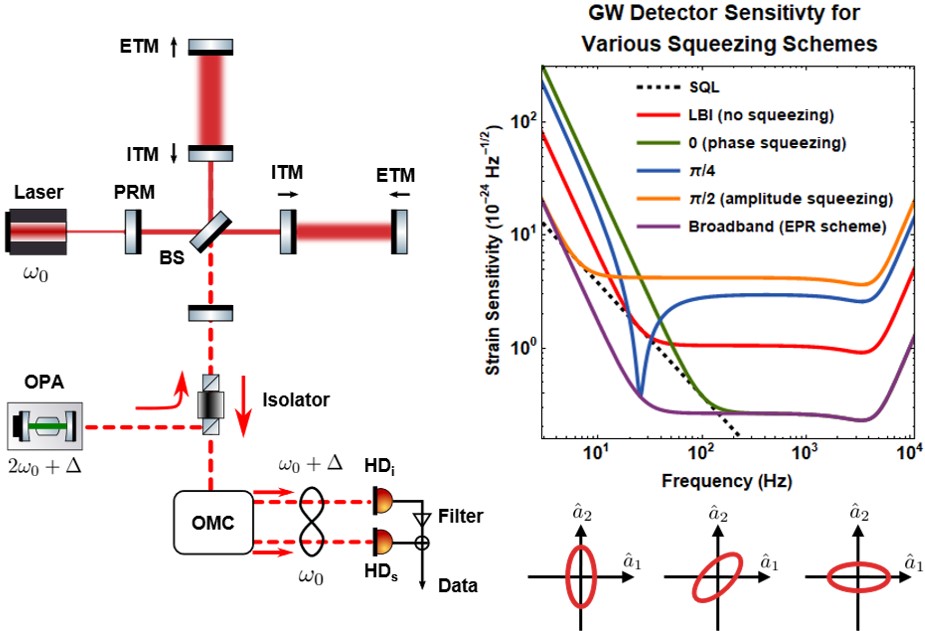}
    \caption{LBI setup and sensitivity curves for various squeezing schemes. A fixed squeezing angle only surpasses the SQL (black dotted line) over a narrow frequency band. Our broadband squeezing scheme (purple curve) is achieved by rotating the noise ellipse in a frequency-dependent way as shown below the plot. Acronyms used are: end test mirror (ETM), input test mirror (ITM), power recycling mirror (PRM), signal recycling mirror (SRM), and output mode cleaner (OMC).}
\label{fig:SetupAndSensitivity}
\end{figure}

The two primary noise sources in GW interferometers are radiation-pressure noise and photon shot noise. The former is due to the quantum fluctuations in the amplitude that cause random fluctuations in the radiation pressure force on the mirrors. The latter is due to the uncertainty in the phase of the light which manifests itself as the statistical arrival time of photons. These noise sources are modelled as entering through the dark port of the interferometer and coupling to the coherent laser light. The detector sensitivity achieved when these uncorrelated (random) fluctuations enter the interferometer is called the \textbf{Standard Quantum Limit} (SQL). It has been shown that this limit could be surpassed if a squeezed vacuum was injected into the interferometer's dark port instead \cite{CavesQMNoiseInIFO,schnabel_2017}. Depending on the squeezing angle of the injected squeezed vacuum (see Fig.\,\ref{fig:SetupAndSensitivity}), we can decrease the amplitude or phase fluctuations that limit the detector sensitivity. 
Amplitude and phase quadratures are conjugate variables (like position and momentum), thus the Heisenberg Uncertainty Principle states that the product of their uncertainties must be greater than some constant. Thus, by decreasing phase fluctuations, we suffer an increase in amplitude fluctuations. This would not be a problem if at low gravitational wave frequencies, the mirror suspension systems did not have mechanical resonances that amplify these fluctuations and make radiation pressure noise the limiting noise source. Put simply, at low frequencies we need amplitude-squeezed vacuum injection. Once far away from these resonances (at higher frequencies), the detector is then limited by shot noise and thus phase-squeezed vacuum is needed. It has been known for some time that frequency-dependent squeezing would allow one to surpass the SQL over all frequencies \cite{KLMTV}. These proposals require additional low-loss filter cavities. It was shown recently that an alternative approach to achieving frequency-dependent squeezing without additional cavities is using EPR-entangled signal and idler beams (different frequency components in conventional squeezed light source)\,\cite{NaturePhysicsEPR}, which has been demonstrated in proof-of-principle experiments\,\cite{Jan2019, Yap2019}. In this paper, we present a systematic way of finding the working points for this broadband squeezing scheme in LBIs. We also show that in LBIs, the EPR scheme achieves nearly perfect ellipse rotation as compared to 4km interferometers which have appreciable error. Finally, we show that an approximation for the opto-mechanical coupling constant in the 4km case can break down for longer baselines.
\section{Theory}
\subsection{EPR Entanglement in Squeezed-light Source}
\begin{figure}[b]
  \includegraphics[width=0.95\columnwidth]{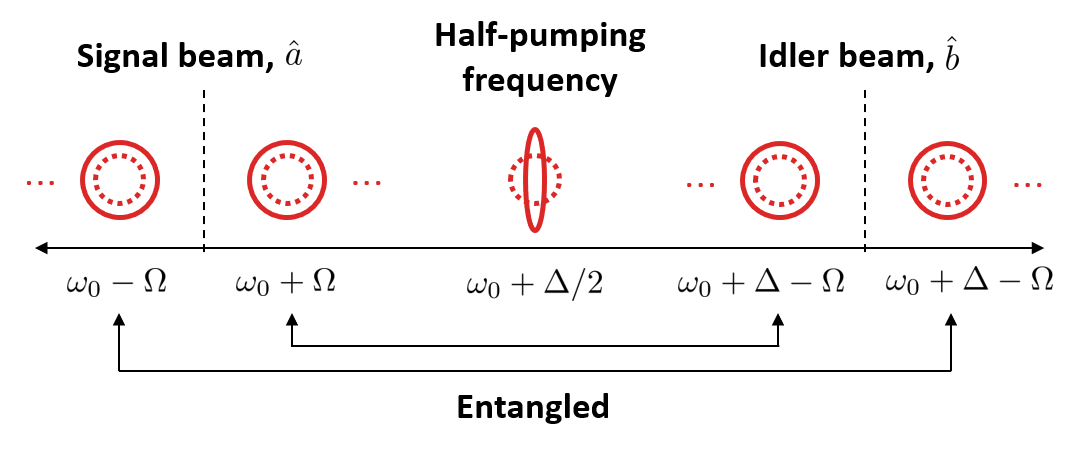}
    \caption{Visualization of the frequency-mode entanglement in the pumped OPA in our proposed scheme.}
\label{fig:DetunedOPA}
\end{figure}

In this section, we will illustrate the EPR entanglment generated in the squeezed-light source, which consists of
 a non-degenerate optical parametric amplifier with a nonlinear electric susceptibility ($\chi^{(2)}$ in our case). Such a device takes in two modes and a pump beam (energy source) and produces two amplified modes which  we call the signal and idler beams. In frequency space, we can visualize the OPA taking in uncorrelated sidebands and entangling (correlating) them. In fact, any frequency modes $\omega_1$, $\omega_2$ within the squeezing bandwidth that suffice $\omega_p = \omega_1 + \omega_2$
will be entangled with each other. In our proposed scheme \cite{EPRwithFINESSE, NaturePhysicsEPR}, we detune the pump field by an amount $\Delta$ (of order MHz) such that $\omega_p = 2\omega_0 + \Delta$, where $\omega_0$ is the interferometer carrier frequency. This creates correlated sidebands around the frequencies $\omega_0$ and $\omega_0+\Delta$ (Fig.\,\ref{fig:DetunedOPA}). 
The corresponding amplitude and phase quadratures defined with respect to $\omega_0$ and $\omega_0+\Delta$ are therefore entangled\,\cite{ReidEPR,EPR-CV}. 

In Caves and Schumaker's two-photon formalism \cite{TwoPhotonFormalism}, the amplitude and phase quadratures are written in terms of sidebands.
\begin{align}
    \hat{a}_1 (\Omega) &= \frac{\hat{a}(\omega_0+\Omega)+\hat{a}^{\dagger}(\omega_0-\Omega)}{\sqrt{2}}\,,\\ \hat{a}_2 (\Omega) &= \frac{\hat{a}(\omega_0+\Omega)-\hat{a}^{\dagger}(\omega_0-\Omega)}{i\sqrt{2}}\,,\\ 
    \hat{b}_1 (\Omega) &= \frac{\hat{b}(\omega_0+\Delta+\Omega)+\hat{b}^{\dagger}(\omega_0+\Delta-\Omega)}{\sqrt{2}}\,,\\
   \hat{b}_2 (\Omega) &= \frac{\hat{b}(\omega_0+\Delta+\Omega)-\hat{b}^{\dagger}(\omega_0+\Delta-\Omega)}{i\sqrt{2}}\,.
\end{align}
The general quadratures for the signal and idler beams can then be written as
\begin{align}
    \hat{a}_{\theta} &= \hat{a}_1 \cos{\theta} + \hat{a}_2 \sin{\theta}\,, \\ 
     \hat{b}_{\theta} &= \hat{b}_1 \cos{\theta} + \hat{b}_2 \sin{\theta}\,.
\end{align}
In the high squeezing regime, the fluctuations in the joint quadratures, $\hat{a}_1 - \hat{b}_1$ and $\hat{a}_2+\hat{b}_2$, are simultaneously well below the vacuum level. This is the experimental signature of the EPR entanglement. This does not violate Heisenberg's Uncertainty Principle because $[\hat{a}_1 - \hat{b}_1,\hat{a}_2+\hat{b}_2]=0$. In analogy to the original EPR paper \cite{EPR1935,ReidEPR}, $\hat{a}_{-\theta}$ is maximally correlated with $\hat{b}_{\theta}$. This correlation allows us to reduce our uncertainty in $\hat{a}_{-\theta}$ by making a measurement on $\hat{b}_{\theta}$. This key theoretical result that enables our conditional squeezing scheme has been realized experimentally recently \cite{ExperimentalBroadband}.
\subsection{Interferometer as Detector and Filter}
The signal and idler beams enter the dark port of the interferometer and couple to the laser that enters the bright port. The interferometer  we consider has both signal and power recycling cavities to increase sensitivity as shown in Fig. \ref{fig:SetupAndSensitivity}.
To the signal beam, the interferometer looks like a resonant cavity. The input-output relation for the phase quadrature of the signal beam (which will contain our GW signal) is \cite{KLMTV}
\begin{equation}\label{eq:SignalIO}
    \hat{A}_2 = e^{2i\beta}(\hat{a}_2 - \mathcal{K} \hat{a}_1) + \sqrt{2 \mathcal{K}} \frac{h}{h_{\text{SQL}}} e^{i\beta}\,,
\end{equation}
where $h_{\text{SQL}}=\sqrt{8\hbar/(m \Omega^2 L_{\rm arm}^2)}$ is the square root of the SQL, $\beta$ is a phase shift given as $\beta\equiv \arctan{\Omega/\gamma}$ with $\gamma$ being the detection bandwidth, and $\mathcal{K}$ is the optomechanical coupling constant that determines the coupling between the light and the interferometer mirrors. For current GW detectors, the signal-recycling cavity length is of the order of 10 meters and the SRM transmission is quite high (tens of percent). In this case, we can effectively view the signal-recycling cavity (SRC) formed by SRM and ITM as a compound mirror by ignoring the propagation phase $\Omega L_{\rm SRC}/c$ picked up by the sidebands, as first mentioned by Buonanno and Chen\,\cite{ScalingLaw}. When the SRC is tuned, the corresponding optomechanical coupling constant $\cal K$ is the same as given by Kimble {\it et al.}\,\cite{KLMTV}:
\begin{equation}
    \mathcal{K}_{\text{KLMTV}}=\frac{32 \omega_0 P_{\text{arm}}}{mL_{\rm arm}^2 \Omega^2 (\Omega^2 + \gamma^2)}\,,
\end{equation}
where $P_{\rm arm}$ is the arm cavity power, $m$ is the mirror mass and $\gamma$ is equal to $ c T_{\rm SRC}/(4 L_{\rm arm})$ with $T_{\rm SRC}$ being the effective power transmission of the SRC when viewed as a compound mirror. 
 
For LBIs, the definition of such a coupling constant can differ form that given by Kimble {it et al}. This is because the SRC length is longer and also the SRM transmission becomes comparable to that of ITM in order to broaden the detection bandwidth in the resonant-sideband-extraction mode, i.e., the effective transmissivity $T_{\rm SRC}$ of SRC approaches 1. The approximation for defining $\gamma$ applied in Ref.\,\cite{ScalingLaw}, which assumes $T_{\rm SRC}\ll 1$, starts to break down. We also need to take into account the frequency dependent propagation phase of the sidebands, which leads to the following expression for the coupling constant constant\,\cite{martynov2019}: 
\begin{equation}\label{Kappa_LBI}
    \mathcal{K}_{\text{LBIs
    }}=\frac{2 h^2_{\text{SQL}} L_{\rm arm} \omega_0 P_{\text{arm}} \gamma_s \omega^2_s}{\hbar c [\gamma_s^2 \Omega^2 + (\Omega^2-\omega^2_s)^2]}\,.
\end{equation}
Here $L_{\text{arm}}$ is the interferometer arm length;  $\omega_s$ is a resonant frequency that arises from the coupling between the signal recycling and arm cavities. The frequency and bandwidth for such a resonance are given by
\begin{equation}
    \omega_s = \frac{c  T_{\text{ITM}}}{2\sqrt{L_{\text{arm}}L_{\text{SRC}}}}\,,\quad
    \gamma_s = \frac{c T_{\text{SRM}}}{4 L_{\text{SRC}}}\,.
\end{equation}

\begin{figure}[t]
  \includegraphics[width=.9\columnwidth]{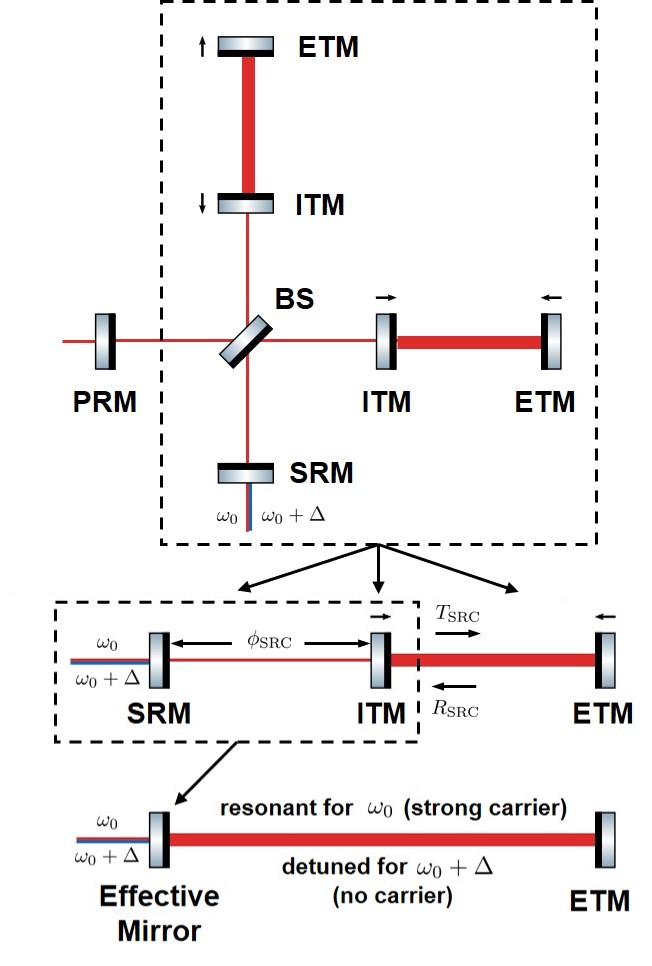}
    \caption{The signal recycling interferometer can be mapped to a three-mirror cavity. The signal recycling cavity can then be mapped into a single mirror with an effective transmissivities and reflectivities \cite{ScalingLaw}. This final, two-mirror cavity is resonant for the signal beam (at $\omega_0$) but detuned for the idler beam (at $\omega_0+\Delta$), thus the idler simply experiences a frequency-dependent ellipse rotation. This allows us to use the interferometer itself as a filter cavity.
    The single-trip propagation phase $\phi_{\rm SRC}$ is equal to integer number of $\pi$ for the carrier in the resonant-sideband-extraction mode, and is equal to some specific number for the idler, as explained in the text.}
\label{fig:Mapping}
\end{figure}

The idler beam, however, is far away from the carrier frequency and does not produce noticeable radiation pressure effect on the test mass. As such, it sees the interferometer as a simple detuned cavity as shown in Fig.\,\ref{fig:Mapping}, the same as done in Ref.\,\cite{ScalingLaw}. 
As such, the idler beam simply experiences a frequency dependent ellipse rotation. This can be seen in the idler input-output relation which is given as
\begin{equation}\label{eq:IdlerIO}
    \hat{B}_2 = e^{i\alpha}(-\hat{b}_1 \sin{\Phi_{\text{rot}}}+\hat{b}_2 \cos{\Phi_{\text{rot}}})\,
\end{equation}
where $\alpha$ is an unimportant overall phase and the
the important rotation angle $\Phi_{\rm rot}$
is given by\,\cite{KLMTV, EPRwithFINESSE}:
\begin{equation}\label{eq:ApproxRot}
\Phi_{\text{rot}} = \arctan{\bigg(\frac{\Omega+\delta_f}{\gamma_f}\bigg)}+\arctan{\bigg(\frac{-\Omega+\delta_f}{\gamma_f}\bigg)}\,.
\end{equation}
Here $\delta_f$ and $\gamma_f$ are the effective detuning and bandwidth of the interferomemeter with respect to the idler beam. They are defined through
\begin{align}\label{resonance}
2(\omega_{\rm idler}+\delta_f) (L_{\rm arm}/c)&+ \arg(r^{\rm idler}_{\rm SRC})= 2n\pi\,,\\
\gamma_f &\equiv c |t^{\rm idler}_{\rm SRC}|^2/
(4L_{\rm arm})\,, \label{eq:approxgammaf}
\end{align}
where $\omega_{\rm idler} = \omega_0 +\Delta$, 
$n$ is an integer number, $r^{\rm idler}_{\rm SRC}$ and 
$r^{\rm idler}_{\rm SRC}$ are the effective amplitude 
reflectivity and transmissivity of the SRC for the idler beam:
\begin{align}
    r_{\rm SRC}^{\rm idler}&= \sqrt{R_{\rm ITM}} +\frac{T_{\rm ITM}\sqrt{R_{\rm SRM}}}{1-\sqrt{R_{\rm ITM}R_{\rm SRM}} e^{2i\phi_{\rm SRC}^{\rm idler}}}\,,\\
    t_{\rm SRC}^{\rm idler}&=\frac{\sqrt{T_{\text{SRM}}T_{\text{ITM}}}e^{i\phi^{\rm idler}_{\text{SRC}}}}{1-\sqrt{R_{\text{ITM}}R_{\text{SRM}}}e^{2i\phi^{\rm idler}_{\text{SRC}}}}\,. \label{tSRC}
\end{align}
The phase $\phi_{\rm SRC}^{\rm ider} = \Delta L_{\rm SRC}/c $ by assuming $\omega_0 L_{\rm SRC}/c$ equal to 
integer number of $\pi$ as in the resonant-sideband-extraction mode. 
Note that the issue 
of the compound-mirror approximation for the carrier mentioned earlier does not occur for the idler beam. Because $\Delta \gg \Omega$, the sideband propagation phase inside SRC can be ignored, and also 
the effective SRC transmissivity for the idler $T_{\rm SRC}^{\rm idler} = |t^{\rm idler}_{\rm SRC}|^2$ is much smaller than 1, which makes $\gamma_f$ properly defined. 

The rotation angle $\Phi_{\text{rot}}$ needs to be equal to $\arctan{\mathcal{K}}$ to achieve the required frequency-dependent squeezing. This usually cannot be 
realised exactly with a single cavity, and two cavities are required. Indeed, for LIGO implementation of such an idea\,\cite{NaturePhysicsEPR}, the rotation in the intermediate frequency does deviate from the ideal one
by a noticeable amount. As we will see, for LBIs, the broadband operation mode can make such a deviation negligible, because the transition frequency from the radiation-pressure-noise dominant regime to the shot-noise dominant regime is much lower than the detection bandwidth; a single filter cavity is close to be sufficient and ideal\,\cite{Khalili}. The corresponding required detuning and bandwidth for given 
Eq.\,\eqref{Kappa_LBI} and $\omega_s\gg \Omega$,
following a similar 
derivation as Refs.\,\cite{Khalili,NaturePhysicsEPR}:
\begin{align}\label{eq:gammaf}
    \gamma_f = \sqrt{\frac{\Omega^2 \mathcal{K}_{\rm LBIs}}{2}}\Bigg|_{\omega_s\gg \Omega} & \approx  \sqrt{\frac{4 \omega_0 P_{\text{arm}}T_{\rm SRM}}{m c^2 T^2_{\text{ITM}}}}\,,\\
    \delta_f &= -\gamma_f\,.
    \label{eq:deltaf}
\end{align}
From Eq.\,(46) in \cite{KLMTV}, one can show that the sensitivity of the interferometer with imperfect rotation angle is 
\begin{equation}\label{eq:Sh_loss}
     S_h \approx \frac{h^2_{\text{SQL}}}{2 \cosh{2r}}\bigg(\mathcal{K} + \frac{1}{\mathcal{K}}\bigg) + \frac{h^2_{\text{SQL}}}{2}\frac{\sinh^2{2r}}{\cosh{2r}}\bigg(\mathcal{K} + \frac{1}{\mathcal{K}}\bigg)\delta \Phi^2
\end{equation}
where $r$ is the squeezing factor and $\delta \Phi = \Phi_{\text{rot}}-\arctan{\cal K}<< 1$. The first term in Eq.\,\eqref{eq:Sh_loss} is the sensitivity when the rotational angle is realized exactly and the second term is the degradation in sensitivity due to error in the rotational angle. In the case of a 15dB squeezing injection as considered in \cite{NaturePhysicsEPR}, $r=1.73$, so the ratio of the correction term to the exact term is $\approx 249 \delta \Phi^2$. So, if we want to keep the relative correction to less than 10\%, we will need the error in the rotation angle $\delta \Phi < 0.02$ rad (i.e $249 (0.02\text{ rad})^2 \times 100\% = 9.96\% < 10\%$).
So, as long as the proposed scheme keeps the overall error in the rotation angle to less than $0.02$ rad, we will suffer no more than a $10\%$ degradation in noise reduction. This requirement turns out to be easily satisfied in the broadband detection mode of LBIs due to the reason mentioned earlier. 

\section{Numerical Results}

In this section, we will show the systematic approach to finding the working points for implementing this idea. The procedure was outlined in Ref.\,\cite{NaturePhysicsEPR}, which showed one working points out of many for LIGO parameters. The tunable parameters are the detuning frequency of the pump $\Delta$, the small change of the arm length $\delta L_{\rm arm}$ and the SRC length $\delta L_{\rm SRC}$
with respect to their macroscopic value. We show how the relevant domain for the various tunable parameters in our scheme were derived. Then, we present the result of our search for solutions to the resonance condition within these bounds. 

For illustration, we choose the detector parameters on the scale similar to the Einstein Telescope. To highlight that the EPR squeezing scheme is not restricted to one particular set of detector design parameters, we allow the macroscopic arm length $L_{\rm arm}$ and $L_{\rm SRC}$ to vary from the nominal values outlined in the Einstein Telescope design study \cite{ETDesign}. The detector  parameters are outlined in the following table.

\bgroup
\def\arraystretch{1.2}

\begin{center}
\begin{tabular}{ |c|c|c|c| } 
\hline
Parameters & Name & Value \\
\hline
$L_{\text{arm}}$& arm cavity length &[9995, 10005]\,m \\ 
$L_{\text{SRC}}$& signal recycling cavity length & [100, 200]\,m\\
$m$ & mirror mass &150\,kg \\
$T_{\text{ITM}}$&ITM power transmissivity & 0.04 \\
$T_{\text{SRM}}$&SRM power transmissivity & 0.04 \\
$P_{\text{arm}}$&intra-cavity power & 3 MW \\
$r$ & squeezing parameter & 1.73 (15\,dB)\\
\hline
\end{tabular}
\end{center}

To start, we equate Eq. \eqref{eq:approxgammaf} with Eq. \eqref{eq:gammaf} and solve for  $\phi_{\text{SRC}}$. This is the exact phase accumulated by the idler after one round trip in the SRC, so we denote it $\phi^{\text{exact}}_{\text{SRC}}$.
 Next, for the effective cavity to have a detuning frequency satisfying Eq.\,\eqref{eq:deltaf}:  $\delta_f=-\gamma_f$, we tune the idler detuning $\Delta$, $\delta L_{\rm arm}$ and $\delta L_{\rm SRC}$ to find solutions to Eq. \eqref{resonance}. The idler detuning $\Delta$ has to be in the low MHz regime because if it was lower it would interfere with the carrier, but if it were too high, electronics would not work optimally. The allowable range is taken to be 
\begin{equation}
    \frac{\Delta}{2\pi} \in [5,50] \,\text{MHz}
\end{equation}
Since we want to keep $\gamma_f$ 
 fixed while we tune $\Delta$ to make the resonance condition Eq.\,\eqref{resonance} satisfied for $\delta_f = -\gamma_f$, we can only alter $\Delta$ by integer numbers $n$ of the free spectral range of the SRC, namely $\Delta = (\phi^{\text{exact}}_{\text{SRC}} + n \pi) c/L_{\rm SRC}$.   
The minimum allowed detuning is 5\,MHz and this will occur when $L_{\text{SRC}}$ is at its maximum, i.e. $200$\,m. This will correspond to the minimum allowed $n$. Similarly, the max detuning occurs to the minimum SRC length and the max $n$. We find the relevant values of $n$ to be
\begin{equation}
    n \in [7,33]
\end{equation}

 We found that the overall rotation angle $\Phi_{\rm rot}$ is not very sensitive to changes in the SRC phase. It is acceptable to have $\phi_{\text{SRC}}$ slightly deviated from the exact value  $\phi^{\text{exact}}_{\text{SRC}}$. This makes it easier for us to find solutions to our resonance condition Eq.\,\eqref{resonance}. As noted before, we must keep the error in the overall rotation angle to less than 0.02 rad. That is, we need $|\delta \Phi|=|\Phi_{\text{rot}}-\arctan{\cal K}|<0.02 \text{ rad}$ where $ \Phi_{{\text{rot}}}$ is given in Eq.\,\eqref{eq:ApproxRot}. To ensure the error in rotation angle is less than 0.02 rad over the whole positive frequency domain we require
\begin{equation}
    \max\limits_{\Omega} |\Delta \phi_{\text{SRC}} \frac{d\Phi_{\text{rot}}}{d \phi_{\text{SRC}}}|<0.02
\end{equation}
Using the given parameters listed in the table, one finds 
\begin{equation}
   |\Delta \phi_{\text{SRC}}|<0.002\,.
\end{equation}
\bgroup
\def\arraystretch{1.2}

%

To show the working points for given different macroscopic arm length and SRC length, we scan the length by $1$ m step size. Additionally, we sweep $\phi^{\text{approx}}_{\text{SRC}}$
in between $[ \phi^{\text{exact}}_{\text{SRC}}-0.002,\phi^{\text{exact}}_{\text{SRC}}+0.002]$ with a step size of $0.0001 $. There are 1.2 million combinations of these parameters with the given step sizes. We took advantage of the fact that each combination is independent and can thus be checked in parallel. We define the working point as those requires microsopic change of arm length $\Delta L_{\rm arm}$ and SRC length $\delta L_{\rm SRC}$ smaller than 1\,cm. Our search resulted in 3444 working points summarized in  Fig.\,\ref{fig:WorkingPoints}. 
We pick one of the many working points for illustration, and the result sensitivity curve is given by Fig.\,\ref{fig:ApproxNoiseReduction}. The EPR scheme achieves almost the ideal frequency dependent rotation of the squeezing quadrature angle. This is result of us heavily restricting our parameter space to bound the error.
\begin{figure}[ht]
  \includegraphics[width=\columnwidth]{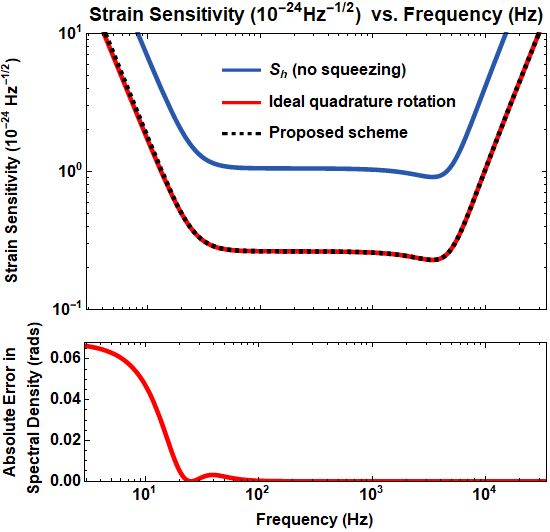}
    \caption{Approximate sensitivity (dotted curve) plotted alongside the sensitivity when ideal ellipse rotation is achieved (red curve). For this plot, we use $L_{\rm arm} = 10003$\,m, $L_{\rm SRC} = 100$\,m, $\Delta/(2\pi) = 1.04$\,MHz and $\phi_{\rm SRC}^{\rm approx} =0.16828$.}
\label{fig:ApproxNoiseReduction}
\end{figure}

\begin{figure*}[ht]
    \centering
    \includegraphics[width=2\columnwidth]{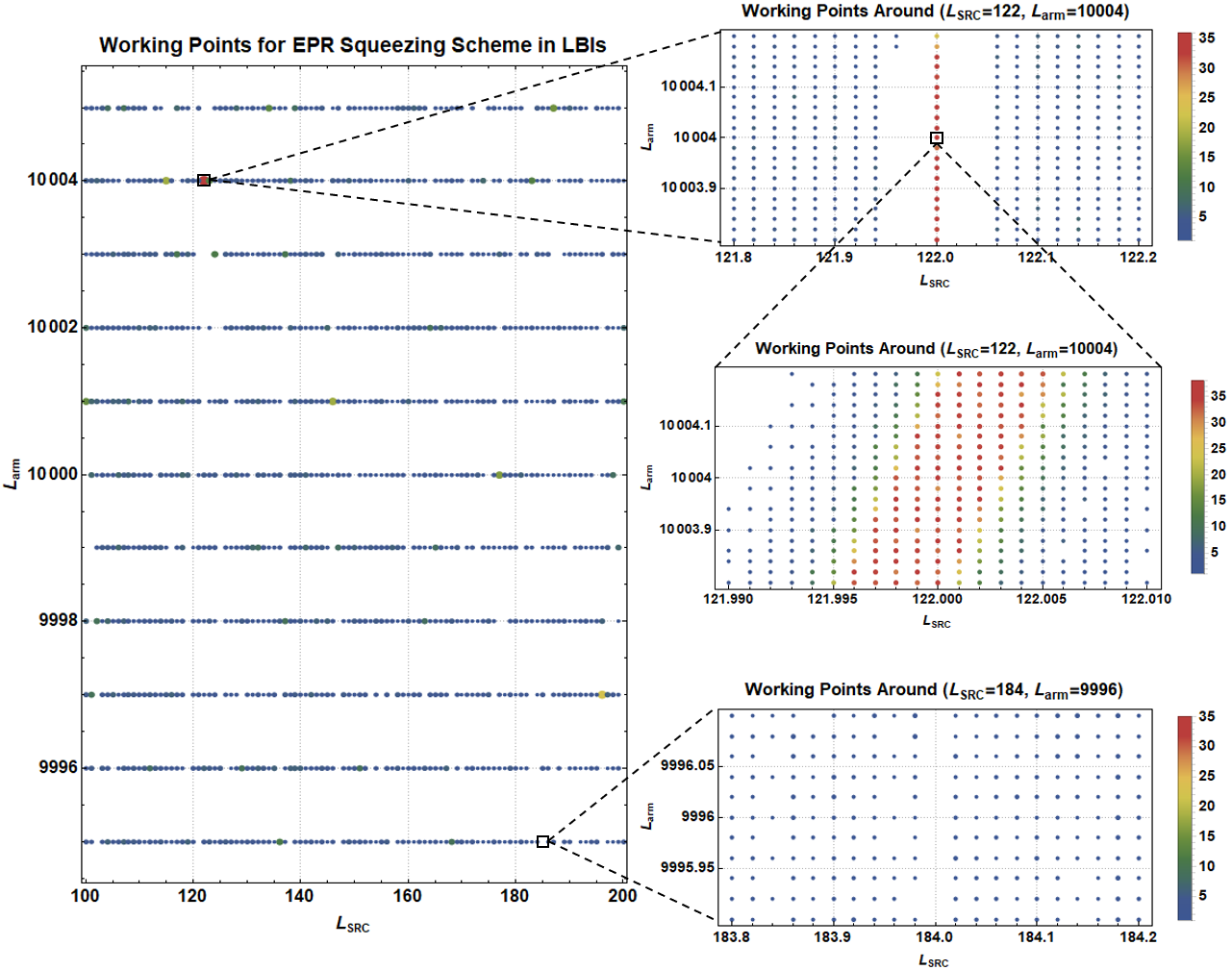}
    \caption{The left panel shows all the working points found at the resolution set by the step size of the arm cavity length and SRC length to be 1 m. The color of the point indicated how many working points exist at a given combination of $L_{\text{SRC}}$ and $L_{\text{arm}}$. The right column of plots are zoomed in considerably. The existence of working points in these plots indicates that with sufficient fine tuning, working points for this squeezing scheme can always be found.}
    \label{fig:WorkingPoints}
\end{figure*}
The step sizes were chosen with computational expense in mind, so the resolution is not particularly high. As such, Fig. \ref{fig:WorkingPoints} shows several ``dead zones'' as well as a couple ``hot points''. The natural question to ask is whether or not these are real or whether they are a byproduct of our numerical precision. Zooming in around two such points, we produced the subfigures on the righthand side of Fig. \ref{fig:WorkingPoints}. In the case of the ``dead zone'', we see that there are actually working points where there appeared to be none. This is promising, as it points to the conclusion that a working point can be found given precise fine-tuning. Similarly, we zoomed in on a ``hot point'' (top right panel of Fig. \ref{fig:WorkingPoints}) and interestingly we still see a line structure where there are as many as 35 working points surrounded by areas that apparently have zero working points. So, to check whether this was a real feature of the system or an issue of numerical precision, we again zoomed in round the ``hot points.'' What we found, once again, is that the ``dead zones'' must be simply due to the numerical precision chosen. With more time or computational power (or both) one could map a relatively smooth landscape of working points for ET. In our case, our goal was to simply show that this EPR-based squeezing scheme is not very sensitive to the actual arm length and SRC length, and that we can always find some working points for given a set of parameters.
\section{Conclusions} 
We have shown that EPR entanglement-based squeezing can be implemented in LBIs. We derived the relevant bounds on the tunable parameters to ensure that our approximate ellipse rotation scheme very nearly matches the exact rotation achievable through the use of external filter cavities. 
The goal of the project was to map the interferometer working points for this squeezing scheme in LBIs like ET and Cosmic Explorer. We accomplished this at a rather low resolution of the parameter space. Zooming in around areas that had very many working points or very few showed that the landscape of working points seems to be quite smooth. In other words, if an area appears to have no working points, it is likely because the step size used to iterate through the parameter space was too low. This is ideal for experimental implementation of such a scheme, for if we cannot fulfill the requirement given the norminal parameters, there should be another working point less than centimeters away. As such, we conclude that EPR-based squeezing is an appealing alternative to other broadband quantum noise reduction schemes that require additional filter cavities.
\section{Acknowledgements} 
We would like to thank P. Jones, and A. Freise for fruitful discussions. J. B. was supported by a Fulbright-University of Birmingham Postgraduate Award during the completion of this research. H. M. has been supported by the UK STFC Ernest Rutherford Fellowship (Grant No.
ST/M005844/11).
\newpage
\bibliography{references}


\end{document}